# Probing metastable space-charge potentials in a wide bandgap semiconductor

Artur Lozovoi[1,*], Harishankar Jayakumar[1,*], Damon Daw[1,2], Ayesha Lakra[1], and Carlos A. Meriles[1,2,†]

[1]Department. of Physics, CUNY-City College of New York, New York, NY 10031, USA.
[2]CUNY-Graduate Center, New York, NY 10016, USA.
[*]Equally contributing authors.
[†]Corresponding author. E-mail: cmeriles@ccny.cuny.edu.

While the study of space charge potentials has a long history, present models are largely based on the notion of steady state equilibrium, ill-suited to describe wide bandgap semiconductors with moderate to low concentrations of defects. Here we build on color centers in diamond both to locally inject carriers into the crystal and probe their evolution as they propagate in the presence of external and internal potentials. We witness the formation of metastable charge patterns whose shape — and concomitant field — can be engineered through the timing of carrier injection and applied voltages. With the help of previously crafted charge patterns, we unveil a rich interplay between local and extended sources of space charge field, which we then exploit to show space-charge-induced carrier guiding.

Charge transport in semiconductor devices — including solar cells, light emitting diodes and other energy-conversion systems — emerges from a subtle interplay between injected carriers and the type and concentration of dopants and traps present in the sample. Given the intimate connection between electric field and carrier propagation, the formation of space charge potentials is central to gaining a full understanding of transport, even in the limit of defect-free materials as early recognized by Mott and Gurney[1]. More broadly, dopant ionization and charge trapping combine to yield complex space charge distributions within the semiconductor, which, in turn, define the current response $I$ to an externally applied voltage $V$. Describing this response — especially for low and intermediate voltages where systems depart from the Mott-Gurney model — has been the subject of work spanning many decades[2-4]. Nonetheless, a general model that correctly reproduces experimental observations without invoking ad-hoc heterogeneity in the distribution of intra-bandgap states has been formulated only recently[5].

Implicitly or explicitly, much of the present formalism rests on the assumption that, after a sufficiently long time, the system reaches a 'steady state' where an effective Fermi level can be defined. In the same vein, the sample is conjectured to start from a thermodynamic equilibrium defining, e.g., the fraction of ionized donors or charged traps. It is precisely the change relative to these equilibrium configurations that governs the formation of space charge potentials in the presence of externally applied fields. While this framework has undeniably proven useful, it is not uncommon, however, to find situations where defects adopt metastable charge states featuring virtually infinite lifetimes. This is often the case in wide-bandgap semiconductors hosting deep donors and acceptors, whose charge can be locally altered, e.g., with the help of optical pulses[6,7]. In this regime, the notion of thermodynamic equilibrium throughout the semiconductor, with or without external fields, breaks down. Further, experimental approaches sensitive to spatial averages — the case for $I$-$V$ measurements[8] — or selectively applicable to surfaces — such as scanning-probe-assisted methods[9,10] — become inadequate, and strategies capable of providing local space charge information must instead be sought.

As a first step to implementing these ideas, here we use confocal fluorescence microscopy to visualize the dynamics of space charge formation in diamond. Specifically, we exploit the nitrogen-vacancy (NV) color center as a local probe to reveal charge carrier dynamics in the presence of externally applied voltages. Combining local optical excitation and fluorescent imaging in areas around the point of illumination, we induce and observe the formation of spatial NV fluorescence patterns, whose shapes — themselves a function of the applied electric potentials — allow us to capture the genesis and spatial distribution of space charge fields. From our experiments, we conclude these fields can reach values comparable to those applied externally (here exceeding $10^6$ V/m). Further, we use time-resolved potentials and optical excitation to show that space charge fields profoundly depend on the preparation protocol, an observation we subsequently build upon to confine and guide carrier propagation.

Figure 1a shows a schematic of our experimental setup[11]: We use a custom-made confocal fluorescence microscope combining green (520 nm) and orange (594 nm) lasers, whose timing we control with the help of on/off logic pulses; to reconstruct a fluorescence image, we scan the orange beam across the sample via a precision galvo system. We use a pair of surface metal electrodes connected to a high-voltage power supply and a home-made, high-speed switch to create variable, time-dependent electric fields in a section of the crystal (Fig. 1b). All our measurements below are carried out at room temperature within the 100 μm gap separating the metal pads; we focus the laser beams 10 μm below the diamond surface, with an axial (i.e., depth) resolution of ~3 μm. Throughout the present experiments we investigate a type 1b synthetic diamond (DDK) grown via chemical vapor deposition; the nitrogen concentration is 0.25

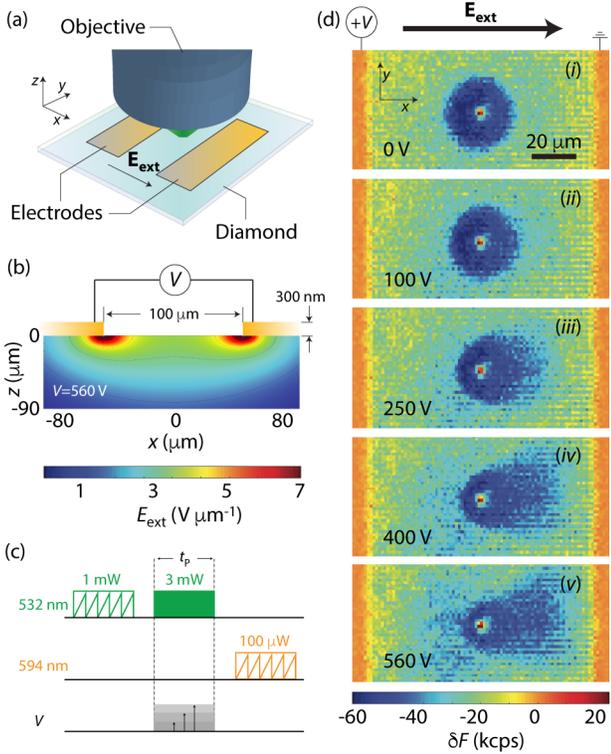

**FIG. 1. Visualizing charge dynamics under constant potentials.** (a) Schematics of the experimental setup. We accelerate photo-generated carriers using a pair of planar electrodes patterned on the diamond surface. (b) Calculated electric field within diamond assuming $V = 560$ V. (c) Experimental protocol; zig-zags indicate laser scans and solid squares correspond to laser parks. (d) Differential NV⁻ fluorescence $\delta F$ as a function of position after a green laser park at the image midpoint for different applied voltages; the laser parking time is $t_P = 1$ min and the arrow denotes the direction of the externally applied electric field **E**. The growing dark tail corresponds to the acceleration of holes towards the right, from high to low potentials.

ppm[12], mostly in the form of neutral substitutional defects or 'P1 centers'.

In the absence of additional sample treatment, only a small fraction of substitutional nitrogen — about one in a hundred[13] — combine with intrinsic vacancies during crystal growth to form NV centers. Unlike P1 centers — which are optically dark — NVs have a bright fluorescence spectrum spanning the near infrared[14]. In the presence of green illumination, two-step one-photon processes interconvert NVs between their neutral and negatively charged states[15], producing in the process equal amounts of carriers of either sign, i.e., ionization of an individual NV⁻ leads to injection of one electron whereas NV⁰ recombination produces a free hole. Therefore, even if their concentration is comparatively low, NVs can act as optically-activated charge pumps able to provide an endless stream of carriers from a predefined (though arbitrary) location in the crystal. Furthermore, because the fluorescence spectra from neutral and negatively-charged NVs overlap only partially, we can make the readout of our microscope selective to only one charge state, which, as we show below, makes it possible to monitor the drift of photo-generated carriers in the applied electric field prior to recapture in a trap.

Figure 1c describes our measurement protocol: To charge initialize the NVs, we first scan the green laser throughout the area between the two electrodes. Because the rate of NV⁻ photo-ionization is lower than the rate of NV⁰ recombination, green illumination leaves the NV ensemble preferentially in the negatively-charged state[15] (~75%). We subsequently park the green beam at the midpoint in the inter-electrode gap, and simultaneously apply a variable voltage to the metal pads. To image the ensuing NV⁻ spatial pattern, we scan the orange laser beam across the area of interest. Although orange excitation can lead to NV⁻ ionization, this process is inefficient at 594 nm[15] implying that a proper selection of the scan parameters (e.g., laser intensity, scanning speed) brings the impact of the readout laser on the NV ensemble to a minimum.

Figure 1d summarizes a series of measurements under variable voltages for a green laser park time of 60 s. Rather than the absolute photon counts, images display the differential fluorescence $\delta F$ obtained upon subtraction of the equivalent fluorescence image in the absence of a green laser park. This procedure allows us to eliminate distortions introduced by uneven excitation and photon collection efficiencies throughout the large field of view required herein (spanning the full inter-electrode gap).

In the absence of externally applied fields, carriers of both signs diffuse radially away from the area under illumination towards the bulk of the crystal where they can be recaptured, both by NVs and other defects. Nitrogen-vacancy centers surrounding the point of green excitation transform efficiently into neutral via trapping of diffusing holes, a process likely amplified by Coulombic attraction[16] towards NV⁻. By contrast, NV⁰ recombination via electron capture is negligible[11], thus leading to the formation of a dark, NV⁰-rich 'halo' concentric with the point of green excitation (case (*i*) in Fig. 1d). As the applied voltage increases, we witness a gradual distortion of the cylindrical symmetry and the formation of a dark 'tail', indicative of hole drift toward the ground electrode.

To gain an understanding of the system dynamics, we first resort to a set of master equations that take into account carrier photo-generation and trapping as well as diffusive transport and electric-field-driven drift, using a set of ionization and capture parameters derived previously[11]. Assuming NV charge initialization via green excitation, Fig. 2a shows the calculated NV charge patterns after a 100-ms-long green park for varying external fields $E_{ext}$. Comparison with the results in Fig. 1d shows the model quickly departs from our observations as it predicts the formation of distinct jet-like patterns, even at fields lower than those applied in our experiments (of order 1 V/μm). This trend changes, however, when the time-varying space charge field created by previously trapped carriers — below referred to as the 'local' contribution to the space charge field $E_{sc}$ — is taken into account throughout the laser park time (right lower image in Fig. 2a, see also Ref. [17]).

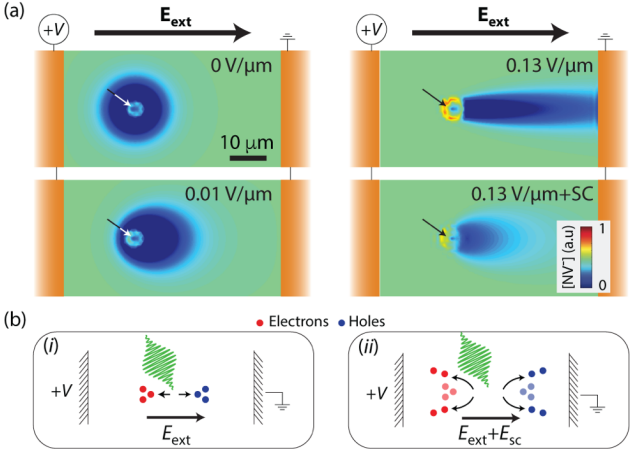

**FIG. 2. Modeling space charge potentials.** (a) Numerical simulation of NV charge patterns for variable field $E_{\text{ext}}$ assuming a diamond with NV (nitrogen) concentration of 2.5 ppb (0.25 ppm). In all calculations, the field is assumed constant throughout the sample. Arrows indicate the point of illumination assuming a beam diameter of 1 μm; the total green laser park time is 20 ms. Space charge (SC) effects are included in the right lower plot, where the induced field is seen to reach 0.075 V/μm. (b) The application of external bias leads to charge separation and thus space charge formation (*i*), subsequently altering the path of incoming carriers (*ii*).

Although the agreement with our observations in Fig. 1 must be deemed qualitative, our modeling shows that the impact of space charge fields on the system dynamics is amplified in the presence of externally applied potentials. Intuitively, the latter can be understood as a consequence of the charge separation taking place at early stages (Fig. 2b), gradually preventing carrier drift toward the electrodes and thus producing the broad, diffusive-like pattern observed experimentally. Unfortunately, using modeling alone to quantify the space charge field in a class 1b diamond is difficult as the type and concentration of all defects at play (including their ionization rates and trapping cross sections) is not fully known[12]. Note, however, that nitrogen impurities — presumably assuming positive, neutral, and negative charge configurations via hole and (multi-) electron capture[17-22] — largely outnumber NVs, implying the latter serve mainly as probes to expose the underlying dynamics.

To more directly gauge the influence of the space charge field, we implement the experiment in Fig. 3a. Similar to the protocol in Fig. 1c, we charge initialize (readout) NV- via a green (orange) laser scan preceding (following) a green laser park. In this case, however, we fraction the park time into a train of $n$ pairs of pulses, each lasting a time $t_{Pj}$, $j = 1,2$, without altering the total park time $\tilde{t}_P = n(t_{P1} + t_{P2})$, which is kept identical to that in Fig. 1. Further, we use a high-speed switch to turn on (off) the electric field during $t_{P1}$ ($t_{P2}$) throughout the pulse train.

Figure 3a shows a collection of NV- differential fluorescence images at increasing voltages. As expected, the charge pattern at zero applied voltage remains unchanged relative to the uninterrupted park (panel (*i*) in Fig. 1d) since the charge dynamics virtually freezes during the dark intervals between laser pulses[6]. As the applied voltage increases, however, the pattern evolves toward a doubly pronged, jet-like shape, dramatically different from that observed in Fig. 1 at comparable voltages. This response can be understood via the conceptual framework of Fig. 2b, namely, the space charge field emerging after $t_{P1}$ causes carriers created during $t_{P2}$ to propagate in a direction opposite to that previously induced by $E_{\text{ext}}$, in the process neutralizing $E_{\text{sc}}$ at the end of the $n$-th cycle. Correspondingly, carriers produced during $t_{P1}$ in the $(n + 1)$-th cycle can propagate farther in the direction of $E_{\text{ext}}$, hence resulting in elongated charge patterns, each resembling those calculated in Fig. 2a. The pattern geometry is indicative of space charge fields comparable to those produced externally. Interestingly, we find that optimally short, voltage-free laser pulses (i.e., $0 < t_{P2} \ll t_{P1}$) are sufficient to allow the formation of narrow, one-directional propagation channels (Fig 3b), suggesting that sources other than the local space charge field — arguably neutralized during $t_{P2}$ — must also be at play in the observations of Fig. 3a. We return to this point later.

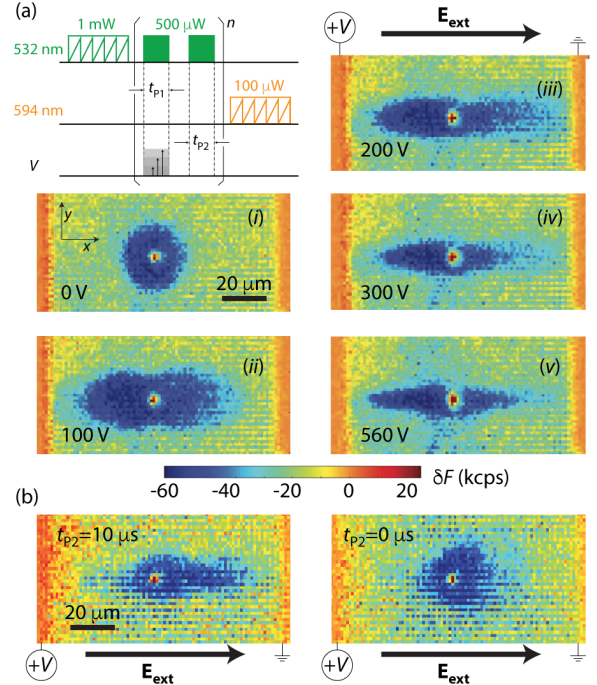

**FIG. 3. Charge dynamics in the presence of pulsed electric fields.** (a) Experimental protocol (top left); the total parking time amounts to $\tilde{t}_P = n(t_{P1} + t_{P2})$. ((*i*) through (*v*)) Differential NV- fluorescence maps for different applied voltages; in these experiments, $t_{P1} = 1$ ms, $t_{P2} = 1$ ms, $\tilde{t}_P = 1$ min, and the arrow denotes the direction of the externally applied electric field. (b) (Left) Same as above but for $t_{P2} = 10$ μs (right) or $t_{P2} = 0$ μs (left) at 420 V. The same color bar applies to both (a) and (b).

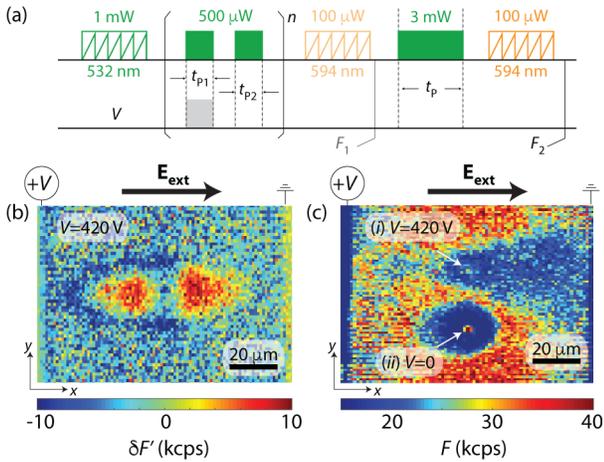

**FIG 4. Probing local and non-local space charge.** (a) Experimental protocol. $F_2$ denotes the final fluorescence image and $F_1$ is the reference image after the first orange scan (optional, shown in a faint trace). (b) Fluorescence image $\delta F' = F_2 - F_1$ upon use of the protocol in (a) for $t_{P1} = t_{P2} = 1$ ms, $n = 5000$, and $\tilde{t}_P = 10$ s. (c) Absolute NV⁻ fluorescence image $F$ upon consecutive, uninterrupted green laser parks at positions (*i*) and (*ii*) (arrows). The voltage is $V = 420$ V during park (*i*) and $V = 0$ V during park (*ii*); all other conditions as in Fig. 1.

Much of the discussion above rests on the implicit idea that local space charge fields combined with trap passivation can steer free carriers in particular directions, a notion we demonstrate with the experiments in Fig. 4a. This time we create a jet-like charge pattern $F_1$ using green laser pulses with alternating voltage (protocol in Fig. 3a), and subsequently apply an uninterrupted, voltage-free green laser park to create a pattern $F_2$. We can then expose the impact of the pre-existing charge distribution on carrier propagation through the differential NV⁻ fluorescence image $\delta F' = F_2 - F_1$. Figure 4b shows the result: The dark contour on the left of the image unmistakably points to carriers propagating along the charge blueprint previously produced during the laser pulse train. Further, the fact that the contour appears exclusively on one side of the pattern indicates a residual, long-lasting space-charge field accelerates holes to the left. Remarkably, we also find a relative growth of NV⁻ fluorescence within the pre-existing blueprint. Unlike the dark contour, the bright pattern features two halves of comparable shape and (differential) intensity suggesting the underlying mechanism is insensitive to the residual space charge field. This observation could indicate the formation of excitons — whose binding energy in diamond reaches 70-80 meV[23] — though additional work will be needed to clarify the dynamics of this process.

Besides this local contribution to the space charge field, we find that other, extended sources play an equally important role. This is shown in Fig. 4c where we park the green laser at two locations (*i*) and (*ii*) for long, uninterrupted intervals (protocol in Fig. 1) differing only in the external electric field, active during the first park but not the second. Remarkably, pattern (*ii*) elongates in a direction opposite to pattern (*i*), thus revealing a non-local field stretching throughout the gap between the electrodes, possibly stemming from carrier traps at the metal-dielectric interface. The amplitude of the tail in pattern (*ii*) suggests this source of space charge is significant. Future work — for example, using the spin of near-electrode NVs as electric field sensors[24] — could prove useful to more quantitatively determine the extent and amplitude of these fields.

In conclusion, transport of photo-generated carriers and charge trapping in adjacent defects leads to the formation of strong, metastable space charge fields, which, in turn, play a dominant role on subsequent carrier propagation. The use of pulsed protocols articulating external voltages and carrier injection allows us to tailor the interplay between carrier propagation and space charge field formation so as to change the carrier range and/or direction of propagation. Further, the charge patterns that emerge can serve themselves as blueprints to guide carriers along trajectories where trapping is strongly suppressed.

Several intriguing questions are in need of further exploration. For example, throughout our measurements we find that local green excitation leads to an overall reduction of NV⁻ population far away from the point of illumination (note the slightly negative value in the background differential fluorescence of Figs. 1 and 3). While we presently cannot rule out the influence of scattered laser light, preliminary observations suggest this effect may instead derive from a form of electric-field insensitive transport, possibly via excitons or exciton-complexes[25,26] formed by electron-hole binding during the laser park. Whether or not this type of process is also responsible for the 'blurring' of pattern (*i*) in Fig. 4c (observed systematically across multiple analogous experiments) is a closely related question. Further examination is thus warranted, especially given the long range of this effect — exceeding several tens of microns — suggestive of trapping-protected propagation.

As a route to probing space charge in insulators, our strategy is complementary to scanning-probe methods[9,10], higher in resolution but restricted to examining surfaces. It also provides an alternative to acoustic- or thermal-stimuli methods[27-30] in that it is light-diffraction-limited and more naturally adapted to the investigation of small defect ensembles. While the present studies focused on diamond, we anticipate comparable behavior in other wide bandgap materials of technological importance. One concrete example is SiC where charge control of SiV and double-vacancy centers has already been demonstrated, and shown to produce long-lived states[31-33]. Another illustration is the family of rare-earth-doped inorganic insulators — of interest for optical information storage[34] due their long-lasting photo-chromism[35] — and lanthanide-hosting nitrides such as GaN[36]. Similar ideas could be applicable to mid- and low-bandgap semiconductors in situations where thermal activation is insufficient to establish equilibrium (e.g., systems under cryogenic conditions). Besides the fundamental aspects, our findings may prove relevant to the operation of various semiconducting devices, especially those combining regions of high and low-doping (i.e., PIN junctions). Combined with photo-current measurements, the present techniques could also find use in the investigation of solar cell operation[37,38], in studies to probe the dynamics of magneto-resistance with sub-micron resolution[39,40], or in the realization of novel quantum technology concepts combining charge and spin[41,42].

**Acknowledgments**. The authors acknowledge support from the National Science Foundation through grant NSF-1914945, and from Research Corporation for Science Advancement through a FRED Award; they also acknowledge access to the facilities and research infrastructure of the NSF CREST IDEALS, grant number NSF-HRD-1547830. This research was supported, in part, under National Science Foundation Grants CNS-0958379, CNS-0855217, ACI-1126113 and the City University of New York High Performance Computing Center at the College of Staten Island.

Supplementary Material for:

# Probing metastable space-charge potentials in a wide bandgap semiconductor

Artur Lozovoi[1,*], Harishankar Jayakumar[1,*], Damon Daw[1,2], Ayesha Lakra[1], and Carlos A. Meriles[1,2,†]

[1]Department. of Physics, CUNY-City College of New York, New York, NY 10031, USA.
[2]CUNY-Graduate Center, New York, NY 10016, USA.
[*]Equally contributing authors.
[†]Corresponding author. E-mail: cmeriles@ccny.cuny.edu.

To describe the diffusion and drift of photo-generated carriers, we solve a set of master equations describing the time evolution of the concentration of holes ($p(\boldsymbol{r},t)$), electrons ($n(\boldsymbol{r},t)$), negatively charged nitrogen vacancies ($Q_-(\boldsymbol{r},t)$), and positively charged substitutional nitrogen impurities ($P_+(\boldsymbol{r},t)$) in the presence of optical excitation and electric fields — both external as well as local-space-charge-induced — as a function of position $\boldsymbol{r}$. We note that although there is evidence that neutral nitrogen can behave as a trap for free electrons[1], there is considerable uncertainty on the electron (hole) capture cross sections of neutral (negatively-charged) nitrogen, hence forcing us, for simplicity, to consider in our calculations only neutral and positively-charged states for nitrogen. The total volume concentration of NV centers and $N_s$ impurities are $Q$ and $P$, respectively. We assume a homogeneous external electric field $\boldsymbol{E}$ applied along the x-axis ($E_x$) and the contribution from the (position-dependent) space charge field $\boldsymbol{E_{SC}}(\boldsymbol{r})$. Then, the complete set of equations is given by:

$$\frac{\partial Q_-}{\partial t} = (\vartheta_0 + \kappa_n n)Q - (\vartheta_0 + \vartheta_- + \kappa_n n + \kappa_p p)Q_- + \Omega[Q(P - P_+) - Q_- P] \,,$$

$$\frac{\partial P_+}{\partial t} = (\vartheta_N + \gamma_p p)P - (\vartheta_N + \gamma_n n + \gamma_p p)P_+ + \Omega[Q(P - P_+) - Q_- P] \,,$$

$$\frac{\partial n}{\partial t} = D_n \nabla^2 n - \mu_n E_x \frac{\partial n}{\partial x} - \mu_n \boldsymbol{\nabla} \cdot (n\boldsymbol{E_{SC}}) + \vartheta_- Q_- + \vartheta_N(P - P_+) - \kappa_n n(Q - Q_-) - \gamma_n n P_+ \,,$$

$$\frac{\partial p}{\partial t} = D_p \nabla^2 p + \mu_p E_x \frac{\partial p}{\partial x} + \mu_p \boldsymbol{\nabla} \cdot (p\boldsymbol{E_{SC}}) + \vartheta_0 (Q - Q_-) - \kappa_p p Q_- - \gamma_p p(P - P_+) \,,$$

where $\vartheta_0$ and $\vartheta_-$ denote the photo-induced NV$^0$→NV$^-$ and NV$^-$→NV$^0$ conversion rates, respectively; $\kappa_p$ and $\kappa_n$ stand for the NV$^-$ hole capture rate and the NV$^0$ electron capture rate. The neutral $N_s$ ionization rate is given by $\vartheta_N$; $\gamma_p$ is the hole capture rate for neutrally charged $N_s$, $\gamma_n$ is the electron capture rate for positively charged $N_s$. Coefficients $D_n$, $\mu_n$, and $D_p$, $\mu_p$ are the electron and hole diffusion constants and mobilities, respectively. The whole set of values used for the calculations is summarized with the corresponding references in Table S1. $\nabla^2$ denotes the Laplace operator in two-dimensions.

We model the space charge fields $\boldsymbol{E_{SC}}(\boldsymbol{r})$ induced at each position by the surrounding charge density using two approximations. First, we consider only electric fields created by substitutional nitrogen in charge states $N_s^0$ and $N_s^+$. This is justified as the concentration of $N_s$ at any time and position is at least 100 times higher than the concentration of NVs, free electrons or holes. Second, we assume that the value of the space charge field is proportional to the gradient of positively charged substitutional nitrogen density at any position, which essentially implies the first-neighbor approximation: $E_{SCx} \propto \frac{\partial P_+}{\partial x}$ and $E_{SCy} \propto \frac{\partial P_+}{\partial y}$. These two approximations significantly decrease the complexity of the system of equations described above.

| $P$ | Substitutional Nitrogen density | 0.25 ppm | |
|---|---|---|---|
| $Q$ | Nitrogen Vacancy density | 2.5 ppb | |
| $Q_-$ | Initial NV⁻ density | 1.8 ppb | |
| $P_0 = P - Q_-$ | Initial N⁰ density | 0.2482 ppm | |
| $\sigma_{Nn}$ | N⁺ electron capture cross section | $3.1 \cdot 10^{-6}$ μm² | [3] |
| $\sigma_{Np}$ | N⁰ hole capture cross section | $1.4 \cdot 10^{-8}$ μm² | [3] |
| $\sigma_{NVp}$ | NV⁻ hole capture cross section | $9 \cdot 10^{-8}$ μm² | [2] |
| $\sigma_{NVe}$ | NV⁰ electron capture cross section | 0 | [2] |
| $I$ | 532 nm illumination power | 1 mW | |
| $I_0$ | Reference 532 nm illumination power | 1 μW | |
| $s$ | Variance of 532 nm laser beam Gaussian power spatial distribution | 1 μm | |
| $\vartheta_N(r)$ | N⁰ photoionization rate under 532 nm illumination | $15 \cdot \frac{I}{I_0} \cdot \exp\left(-\frac{r^2}{s^2}\right)$ Hz | [2] |
| $\vartheta_0(r)$ | NV⁰→NV⁻ recombination rate under 532 nm illumination | $0.0046 \cdot \left(\frac{I}{I_0}\right)^2 \cdot \exp\left(-\frac{r^2}{s^2}\right)$ Hz | [2] |
| $\vartheta_-(r)$ | NV⁻→NV⁰ photoionization rate under 532 nm illumination | $0.0107 \cdot \left(\frac{I}{I_0}\right)^2 \cdot \exp\left(-\frac{r^2}{s^2}\right)$ Hz | [2] |
| $\mu_n$ | Electron mobility in diamond | $2.4 \cdot 10^{11}$ μm²/(V·s) | [3,4] |
| $\mu_p$ | Hole mobility in diamond | $2.1 \cdot 10^{11}$ μm²/(V·s) | [3,4] |
| $D_n = \frac{\mu_n k_B T}{e}$ | Electron diffusion coefficient in diamond | $6.1 \cdot 10^9$ μm²/s | |
| $D_p = \frac{\mu_p k_B T}{e}$ | Hole diffusion coefficient in diamond | $5.3 \cdot 10^9$ μm²/s | |

**Table S1**. List of parameters used in the calculation of NV⁻ distributions formed under diffusion or drift of photo-generated charge carriers presented in Fig. 2 (a) of the main text. The assumed temperature is T=293 K; $k_B = 1.38 \cdot 10^{-23} \frac{J}{K}$ is the Boltzmann constant, and $e = 1.6 \cdot 10^{-19}$ C is the elementary charge. Last column cites references for the values taken from literature.

We solve the equation set using finite element analysis with the help of Matlab's Partial Derivative Equation Toolbox. The results presented in Fig. 2(a) of the main text are obtained for three values of the external electric field $E_x$: $E_x = 0 \frac{V}{\mu m}$, $E_x = 0.01 \frac{V}{\mu m}$, $E_x = 0.13 \frac{V}{\mu m}$. The latter case is also calculated taking into account the presence of the space-charge fields induced by substitutional nitrogen.

**Supplementary references**

[2] S. Dhomkar, P. Zangara, J. Henshaw, C.A. Meriles, "On-demand generation of neutral and negatively-charged silicon-vacancy centers in diamond", *Phys. Rev. Lett.* **120**, 117401 (2018).

[3] L.S. Pan, D.R. Kania, P. Pianetta, O. L. Landen "Carrier density dependent photoconductivity in diamond", *Appl. Phys. Lett.* **57**, 623 (1990).

[4] M. Nesladek, A. Bogdan, W. Deferme, N. Tranchant, P. Bergonzo "Charge transport in high mobility single crystal diamond", *Diam. Relat. Mater.* **17**, 1235 (2008).